\pdfoutput=1

\documentclass[11pt,twoside,a4paper,cmspaper,final,collab]{cms-tdr}
\def\svnVersion{493575P}\def\cmsCernNoTag{CERN-EP-2018-310}\def\cmsCernDate{\today}\def\cmsMessage{Published in the European Physical Journal C as \href{http://dx.doi.org/10.1140/epjc/s10052-019-6788-2}{\doi{10.1140/epjc/s10052-019-6788-2.}}}

\begin{document}\cmsNoteHeader{TOP-17-008}

\hyphenation{had-ron-i-za-tion}
\hyphenation{cal-or-i-me-ter}
\hyphenation{de-vices}
\RCS$Revision: 489823 $
\RCS$HeadURL: svn+ssh://svn.cern.ch/reps/tdr2/papers/TOP-17-008/trunk/TOP-17-008.tex $
\RCS$Id: TOP-17-008.tex 489823 2019-02-22 09:33:11Z jolange $
\newlength\cmsFigWidth
\newlength\cmsTabSkip\setlength{\cmsTabSkip}{1ex}
\ifthenelse{\boolean{cms@external}}{\setlength\cmsFigWidth{0.98\columnwidth}}{\setlength\cmsFigWidth{0.7\textwidth}}

\newcommand{\chisq}{{\ensuremath{\chi^2}}\xspace}

\newcommand{\mtop}{\ensuremath{m_{\cPqt}}\xspace}
\newcommand{\mtfit}{\ensuremath{m_{\cPqt}^\text{fit}}\xspace}
\newcommand{\mtgen}{\ensuremath{m_{\cPqt}^\text{gen}}\xspace}
\newcommand{\mtreco}{\ensuremath{m_{\cPqt}^\text{reco}}\xspace}
\newcommand{\mWreco}{\ensuremath{m_{\PW}^\text{reco}}\xspace}
\newcommand{\Pgof}{\ensuremath{P_\text{gof}}\xspace}
\newcommand{\DRbb}{\ensuremath{\Delta R(\cPqb\cPaqb)}\xspace}
\newcommand{\JSF}{\ensuremath{\text{JSF}}\xspace}

\newcolumntype{d}[1]{D{.}{.}{#1}}
\newcolumntype{L}{>{$}l<{$}}
\newcolumntype{R}{>{$}r<{$}}
\newcolumntype{C}{>{$}c<{$}}

\newcommand{\LH}{\ensuremath{\mathcal{L}}\xspace}
\newcommand{\LHA}{\ensuremath{\LH_\text{A}}\xspace}
\newcommand{\LHL}{\ensuremath{\LH_\text{L}}\xspace}

\newcommand{\statJSF}{\ensuremath{\,\text{(stat+JSF)}}\xspace}
\newcommand{\CRERD}{\ensuremath{\,\text{(CR+ERD)}}\xspace}

\cmsNoteHeader{TOP-17-008}
\title{Measurement of the top quark mass in the all-jets final state at $\sqrt{s}=13\TeV$ and combination with the lepton+jets channel}
\titlerunning{Top quark mass measurement}

\date{\today}

\abstract{
  A top quark mass measurement is performed using $35.9\fbinv$ of LHC proton-proton collision data
  collected with the CMS detector at $\sqrt{s}=13\TeV$.
  The measurement uses the \ttbar{} all-jets final state.
  A kinematic fit is performed to reconstruct the decay of the \ttbar~system and suppress the multijet background.
  Using the ideogram method, the top quark mass (\mtop) is determined, simultaneously constraining an additional
  jet energy scale factor (\JSF).
  The resulting value of $\mtop=172.34\pm0.20\statJSF\pm0.70\syst\GeV$ is in good agreement with previous measurements.
  In addition, a combined measurement that uses the \ttbar{} lepton+jets and all-jets final states is presented, using the same
  mass extraction method, and provides an \mtop measurement of $172.26\pm0.07\statJSF\pm0.61\syst\GeV$.
  This is the first combined \mtop extraction from the lepton+jets and all-jets channels through a single likelihood function.
}

\hypersetup{%
pdfauthor={CMS Collaboration},%
pdftitle={Measurement of the top quark mass in the all-jets final state and the combined lepton+jets and all-jets final states at sqrt(s)=13 TeV},%
pdfsubject={CMS},%
pdfkeywords={CMS, top, quark, mass, hybrid, all, jets, hadronic, fully-hadronic}}

\maketitle

\section{Introduction}

The top quark~\cite{Abe:1995hr,D0:1995jca} is the most massive known fundamental particle and its mass \mtop is an important parameter of the standard model (SM) of particle physics.
Precise measurements of $\mtop$ can be used to test the internal consistency of the SM~\cite{ALEPH:2010aa,Baak:2012kk,Baak:2014ora}
and to search for new physical phenomena.
Since the top quark dominates the higher-order corrections to the
Higgs boson mass, a precise $\mtop$ determination is crucial to put constraints on the stability of the electroweak vacuum~\cite{HiggsStab,HiggsStab1}.

\begin{sloppypar}
At the CERN LHC, top quarks are predominantly produced in quark-antiquark pairs (\ttbar) through the gluon fusion process, and
decay almost exclusively to a bottom quark and a \PW~boson.
Each \ttbar event can be classified through the decays of the \PW~bosons.
Events in the all-jets final state correspond to those that have both $\PW$ bosons decaying further into $\Pq\Paq'$ pairs,
while events in the lepton+jets final state have one \PW~boson decaying to a charged lepton and a neutrino.
\end{sloppypar}

This paper presents a measurement of $\mtop$ obtained in the \ttbar all-jets decay channel
using proton-proton ($\Pp\Pp$) collision data taken in 2016 by the CMS experiment at a center-of-mass energy of $\sqrt{s}=13\TeV$,
corresponding to an integrated luminosity of $35.9\fbinv$.
The two bottom quarks and the four light quarks from the \ttbar decay are all required to be physically separated in the laboratory frame of reference,
and the nominal experimental signature is therefore characterized by six jets in the detector.

Although this final state provides the largest branching fraction of all \ttbar~decays, this measurement of $\mtop$ is particularly challenging,
because of the large background from multijet production.
A kinematic fit of the decay products to the \ttbar hypothesis is therefore employed to
separate signal from background events.

\begin{sloppypar}
The value of \mtop is extracted using the ideogram method~\cite{Abdallah:2008xh,Chatrchyan:2012cz},
which is based on a likelihood function that depends either just on the mass parameter \mtop , or on \mtop combined with an additional jet energy scale factor (JSF).
In the second case, the invariant mass of the two jets associated with the  $\PW  \rightarrow \cPq\cPaq'$
decay serves as an observable to directly estimate the JSF.
\end{sloppypar}

\begin{sloppypar}
Previous measurements in this decay channel have been performed by Tevatron and LHC experiments at lower
center-of-mass energies~\cite{Aaltonen:2014sea,Chatrchyan:2013xza,Khachatryan:2015hba,Aad:2014zea,Aaboud:2017mae}.
The most precise one of these has been obtained by CMS at $\sqrt{s}=8\TeV$,
resulting in a mass of
$\mtop=172.32\pm0.25\statJSF\pm0.59\syst\GeV$.
Combining the results of several measurements using different final states at $\sqrt{s}=7$ and $8\TeV$, ATLAS and CMS reported
values of $\mtop=172.69\pm0.48\GeV$~\cite{Aaboud:2018zbu} and $172.44\pm0.48\GeV$~\cite{Khachatryan:2015hba}, respectively,
while a value of $\mtop=174.30\pm0.65\GeV$ was obtained by combining the Tevatron results~\cite{TevatronElectroweakWorkingGroup:2016lid}.
\end{sloppypar}

\begin{sloppypar}
The top quark
mass has been measured for the first time with $\Pp\Pp$ data at $\sqrt{s} = 13\TeV$,
using the lepton+jets channel~\cite{Sirunyan:2018gqx}, yielding a value of
$\mtop = 172.25\pm0.08\statJSF\pm0.62\syst\GeV$.
A measurement using
both \ttbar all-jets and lepton+jets events is presented here.
This is possible since the two measurements use the same mass extraction method,
so a single likelihood can be used, rather than just combining the two
results statistically.
With this approach, no assumptions on correlations between different uncertainties of the
measurements have to be made.
This is the first report of a combined \mtop measurement in the lepton+jets and all-jets final states using a single likelihood function.
\end{sloppypar}

\section{The CMS detector and event reconstruction}
The central feature of the CMS apparatus is a superconducting solenoid of 6\unit{m} internal diameter, providing a magnetic field of 3.8\unit{T}. Within the solenoid volume are a silicon pixel and strip tracker, a lead tungstate crystal electromagnetic calorimeter (ECAL), and a brass and scintillator hadron calorimeter (HCAL), each composed of a barrel and two endcap sections. Forward calorimeters extend the pseudorapidity ($\eta$) coverage provided by the barrel and endcap detectors. Muons are detected in gas-ionization chambers embedded in the steel flux-return yoke outside the solenoid.

Events of interest are selected using a two-tiered trigger system~\cite{Khachatryan:2016bia}. The first level, composed of custom hardware processors, uses information from the calorimeters and muon detectors to select events within a time interval of 4\mus{}, resulting in a trigger rate of around 100\unit{kHz}. The second level, known as the high-level trigger (HLT), consists of a farm of processors running a version of the full event reconstruction software optimized for fast processing, and reduces the event rate to around 1\unit{kHz} before data storage.

\begin{sloppypar}
The particle-flow (PF) algorithm~\cite{Sirunyan:2017ulk} aims to reconstruct and identify each individual particle in an event, with an optimized combination of information from the various elements of the CMS detector. The energy of photons is obtained from the ECAL measurement. The energy of electrons is determined from a combination of the electron momentum at the primary interaction vertex as determined by the tracker, the energy of the corresponding ECAL cluster, and the energy sum of all bremsstrahlung photons spatially compatible with originating from the electron track. The energy of muons is obtained from the curvature of the corresponding track. The energy of charged hadrons is determined from a combination of their momentum measured in the tracker and the matching ECAL and HCAL energy deposits, corrected for zero-suppression effects and for the response function of the calorimeters to hadronic showers. Finally, the energy of neutral hadrons is obtained from the corresponding corrected ECAL and HCAL energy.
\end{sloppypar}

\begin{sloppypar}
The reconstructed vertex with the largest value of summed physics-object $\pt^2$ is taken to be the primary proton-proton interaction vertex. The physics objects are the jets, clustered using the jet finding algorithm~\cite{Cacciari:2008gp,Cacciari:2011ma} with the tracks assigned to the vertex as inputs, and the associated missing transverse momentum, taken as the negative vector sum of the transverse momentum \pt of those jets.
\end{sloppypar}

Jets are clustered from PF objects using the anti-\kt algorithm with a distance parameter of 0.4~\cite{Cacciari:2005hq,Cacciari:2008gp,Cacciari:2011ma}.
Jet momentum is determined as the vectorial sum of all particle momenta in the jet, and is found from simulation to be within 5 to 10\% of the true momentum over the whole \pt spectrum and detector acceptance. Additional proton-proton interactions within the same or nearby bunch crossings (pileup) can contribute additional tracks and calorimetric energy depositions to the jet momentum. To mitigate this effect, tracks identified to be originating from pileup vertices are discarded, and an offset correction is applied to correct for remaining contributions from neutral hadrons. Jet energy corrections (JECs) are derived from simulation to bring the measured response of jets to that of particle level jets on average. In situ measurements of the momentum balance in dijet, photon+jet, \PZ{}+jet, and multijet events are used to account for any residual differences in the jet energy scale in data and simulation~\cite{Khachatryan:2016kdb}. Additional selection criteria are applied to each jet to remove jets dominated by anomalous contributions from various subdetector components or reconstruction failures~\cite{CMS-PAS-JME-16-003}.

A more detailed description of the CMS detector, together with a definition of the coordinate system used and the relevant kinematic variables, can be found in Ref.~\cite{Chatrchyan:2008zzk}.

\section{Event selection and simulation}\label{sec:evtsel}

Only jets with $\pt>30\GeV$ reconstructed within $\abs{\eta} < 2.4$ are used in the analysis.
For the identification of jets originating from the hadronization of \cPqb{} quarks, the combined secondary vertex
algorithm (CSVv2) \cPqb~tagger is used~\cite{Sirunyan:2017ezt}.
The chosen working point provides an identification efficiency of approximately $50\%$ with
a probability of misidentifying a $\PQu$/$\PQd$/$\PQs$ quark jet or gluon jet as being a bottom jet of approximately $0.1\%$,
and a misidentification probability for $\PQc$ quark jets of $2\%$.
The hadronic activity, used for the event selection, is defined as the scalar \pt sum of all jets in the event,
\begin{linenomath*}\begin{equation*}
  \HT \equiv \sum_{\text{jets}}\pt.
\end{equation*}\end{linenomath*}
Data events are selected using an HLT that requires
the presence of at least six PF jets with $\pt>40\GeV$ and $\HT>450\GeV$.
Additionally, the HLT requires at least one jet to be \cPqb~tagged.

In the offline selection, an event must contain a well reconstructed vertex localized within
24 cm in the $z$ direction and 2\cm in the $x$--$y$ plane around the nominal interaction point.
Selected events are required to contain at least six jets, at least two of which have
to be tagged as \cPqb~jets.
The sixth jet ($\text{jet}_6$), ordered in decreasing \pt, must fulfill
$\pt(\text{jet}_6) > 40\GeV$, and $\HT>450\GeV$ is required.
The two \cPqb~jets must be separated in $\Delta R = \sqrt{\smash[b]{\Delta\phi^2 + \Delta\eta^2}}$ by
$\DRbb > 2.0$.

The \ttbar signal is simulated at an \mtop of $172.5\GeV$ using the
\POWHEG~v2~\cite{Nason:2004rx,Frixione:2007vw,Alioli:2010xd}
matrix-element (ME) generator in next-to-leading order (NLO) perturbative quantum chromodynamics (QCD).
For the parton distribution functions (PDFs), the NNPDF3.0 NLO set~\cite{Ball:2014uwa} is used with the strong coupling constant value of $\alpS=0.118$.
This is one of the first PDF sets to include the total \ttbar cross section measurements from ATLAS and CMS at
$\sqrt{s}=7$ and $8\TeV$ as input.
The parton shower (PS) and hadronization are handled by
\PYTHIA~8.219~\cite{Sjostrand:2007gs} using the \textsc{CUETP8M2T4} tune~\cite{Skands:2014pea, CMS-PAS-TOP-16-021}
and \GEANTfour is used to simulate the response of the CMS detector~\cite{Agostinelli:2002hh}.
The simulated signal sample is normalized to the integrated luminosity of the data sample using
a cross section of
$\sigma_{\ttbar} = 832\unit{pb}$,
calculated at next-to-next-to-leading order in QCD including resummation of next-to-next-to-leading logarithmic soft gluon terms~\cite{Czakon:2011xx}.
In addition to the default sample, six other samples are used assuming
top quark masses of 166.5, 169.5, 171.5, 173.5, 175.5, and  178.5\GeV,
and using the corresponding cross sections.

For simulated events, a trigger emulation is used.
The residual differences in the trigger efficiency between data and simulation are corrected by applying scale factors
to the simulated events.
These are obtained by measuring the trigger efficiency with respect to a reference \HT~trigger
for both data and simulation.
The parameterized ratio as a function of $\pt(\text{jet}_6)$ and \HT is used to reweight
the simulated events.
Additional $\Pp\Pp$ collisions are included in the simulated events.
These are weighted to match the pileup distribution in data.
Finally, corrections to the jet energy scale and resolution, as well as to the \cPqb~tagging efficiency and misidentification rate,
are applied to the simulated events.

\section{Kinematic fit and background estimation}

To improve the resolution of the top quark mass and decrease the background contribution,
a kinematic fit is applied.
It exploits the known topology of the signal events, \ie, pair production of a heavy particle and antiparticle,
each decaying to $\PW\cPqb$ with $\PW\to\cPq\cPaq'$.
The three-momenta of the jets are fitted such that
\begin{linenomath}
\ifthenelse{\boolean{cms@external}}
{
\begin{align*}
  \chisq = \sum_{j\in \text{jets}}&\left[
  \frac{\left({\pt}_j^\text{reco} - {\pt}_j^\text{fit}\right)^2}{\sigma_{{\pt}_j}^2} +
  \frac{\left(\eta_j^\text{reco} - \eta_j^\text{fit}\right)^2}{\sigma_{\eta_j}^2}\right.\\
  &\left.+\frac{\left(\phi_j^\text{reco} - \phi_j^\text{fit}\right)^2}{\sigma_{\phi_j}^2} \right]
\end{align*}
}
{
\begin{equation*}
  \chisq = \sum_{j\in \text{jets}} \left[
  \frac{\left({\pt}_j^\text{reco} - {\pt}_j^\text{fit}\right)^2}{\sigma_{{\pt}_j}^2} +
  \frac{\left(\eta_j^\text{reco} - \eta_j^\text{fit}\right)^2}{\sigma_{\eta_j}^2} +
  \frac{\left(\phi_j^\text{reco} - \phi_j^\text{fit}\right)^2}{\sigma_{\phi_j}^2}\right]
\end{equation*}
}
\end{linenomath}
is minimized, where all jets assigned to the \ttbar decay system are considered.
The labels ``reco'' and ``fit'' denote the components of the originally reconstructed and the fitted jets, respectively,
and the corresponding resolutions are labeled $\sigma_X$.
The minimization is performed,
constraining the invariant mass of the jets assigned to each $\PW$~boson decay to $m_{\PW}=80.4\GeV$.
As an additional constraint, the two top quark candidates are required to have equal invariant masses.

All possible parton-jet assignments are tested using the leading six jets in the event, but
only \cPqb-tagged jets are used as \cPqb{} candidates and equivalent choices (\eg, swapping the two
jets originating from one \PW~boson) are not considered separately.
Of the remaining 12 possibilities, only the assignment yielding the smallest $\chisq$ is used in the following.
The $\chisq$~value can be used as a goodness-of-fit (gof) measure.
For three degrees of freedom, it is translated into a $p$-value of
\begin{linenomath*}\begin{align*}
  \Pgof\equiv
  1-\operatorname{erf} \left(\sqrt{\frac{\chisq}{2}}\right)
  + \sqrt{\frac{2\chisq}{\pi}}\re^{-\chisq/2}.
\end{align*}\end{linenomath*}
Events are required to fulfill $\Pgof>0.1$ for the best assignment.

In simulation, event generator information can be used to validate the assignment of the reconstructed
jets to the top quark decay products.
Events are classified accordingly as \emph{correct} or \emph{wrong} permutations.
A parton-jet assignment is considered correct if the jets can be matched unambiguously to the right partons
within $\Delta R < 0.3$.
Wrong permutations can occur because of a wrong parton-jet assignment, yielding the smallest $\chisq$ or
jets being out of acceptance, not being reconstructed, or failing the identification requirements.

\begin{figure*}[htb]
  \includegraphics[width=.5\linewidth]{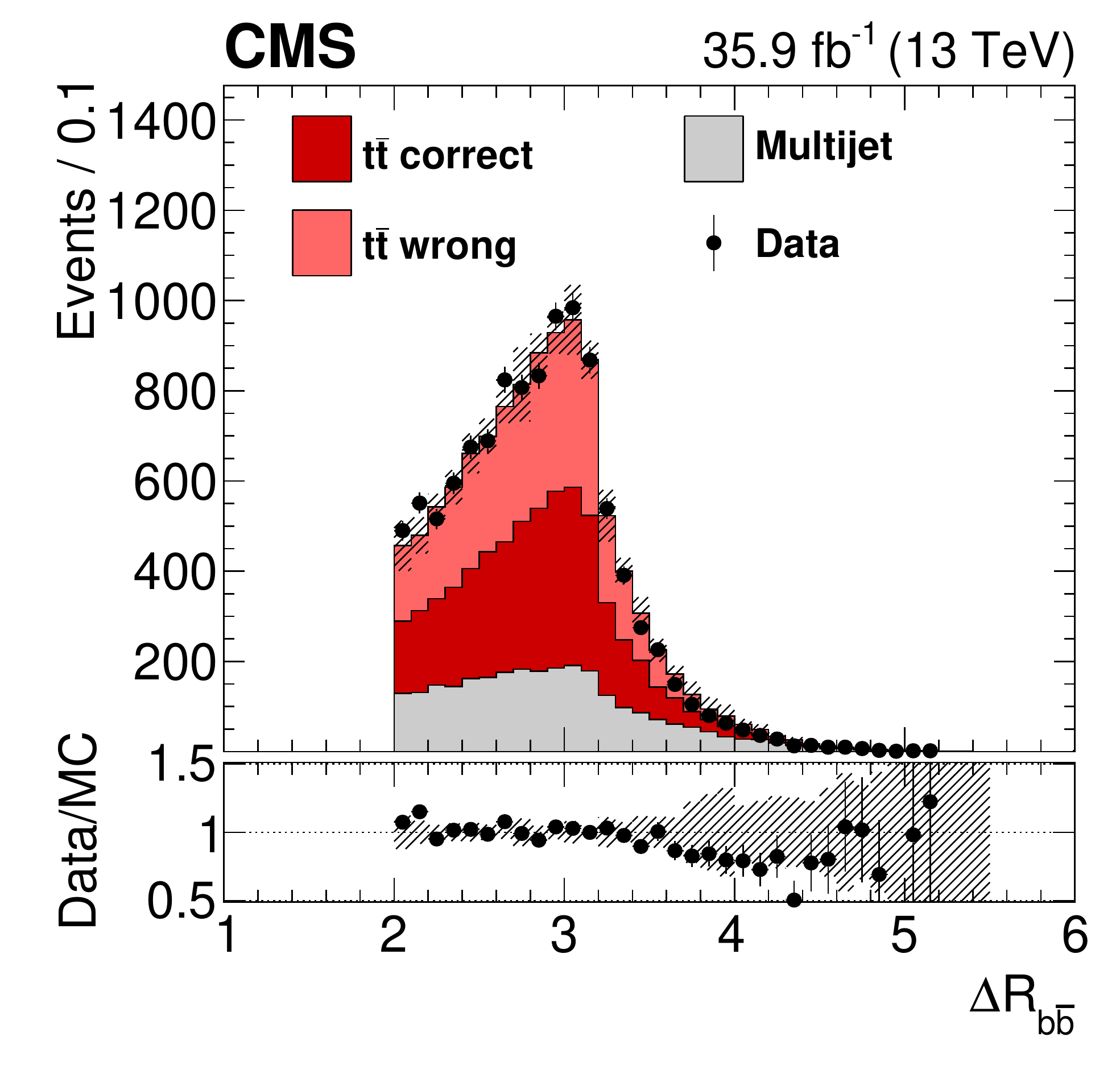}%
  \includegraphics[width=.5\linewidth]{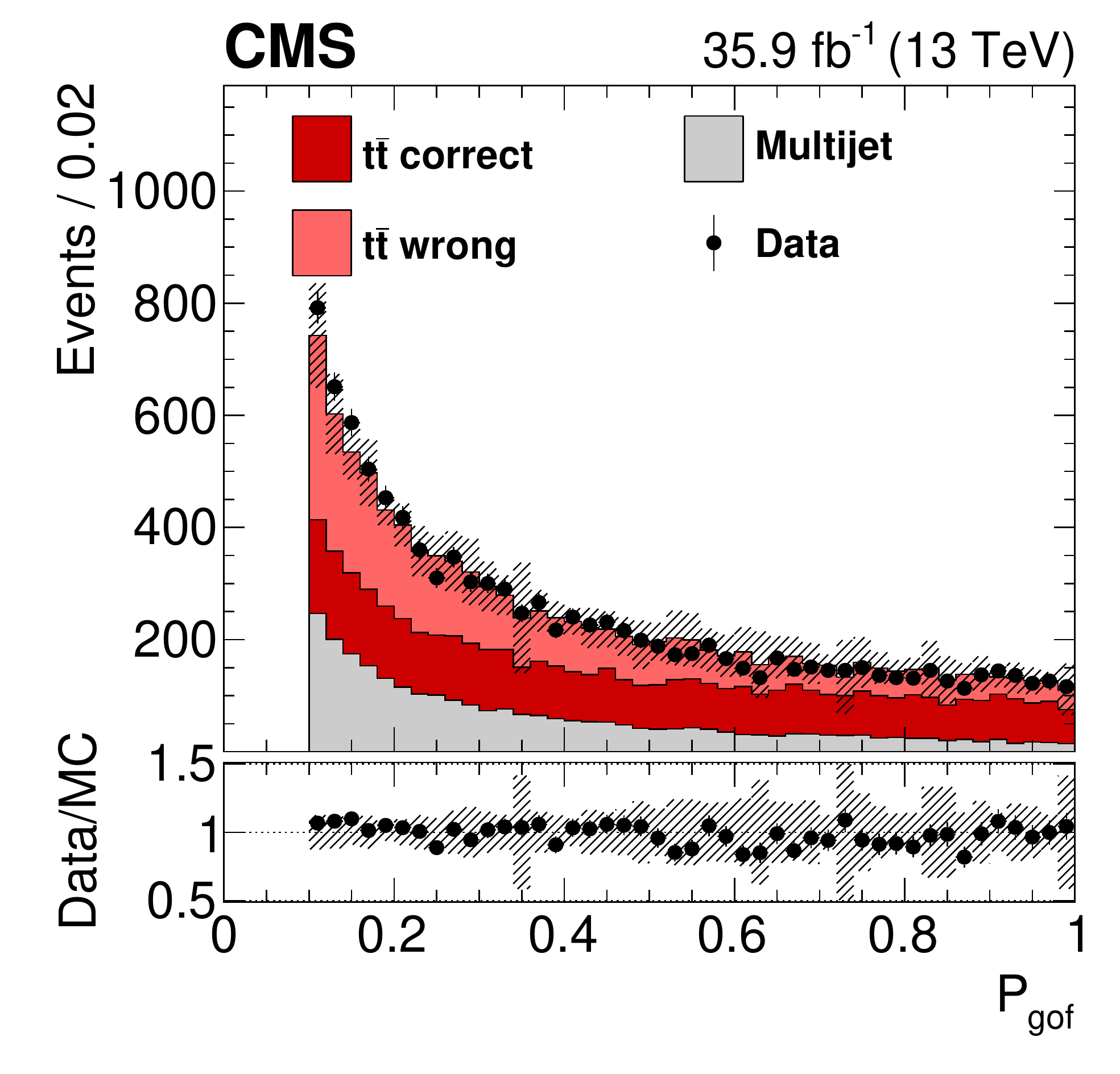}%
  \caption{The \DRbb (left) and \Pgof (right) distributions of data
    compared to simulated signal and the multijet background estimate.
    For each event, the parton-jet assignment yielding the smallest \chisq in the kinematic fit is used.
    The simulated signal events are classified as correct or wrong assignments
    and displayed separately, and the distributions are normalized to the integrated luminosity.
    For the background estimate, the total normalization is given by the difference of observed data events and
    expected signal events.
    The hashed bands represent the total uncertainty in the complete prediction.
    The lower panels show the ratio of data and prediction.
    \label{fig:cp}}
\end{figure*}

The \Pgof distribution is displayed in Fig.~\ref{fig:cp} (right).
Requiring $\Pgof>0.1$ increases the fraction of correct permutations from $6$ to $51\%$.
The fitted top quark mass (\mtfit) is calculated as the invariant mass of the corresponding jets returned by the kinematic fit.
Compared to the mass calculated from the originally reconstructed jets, the
mass resolution is improved from $14.0$ to $8.8\GeV$ for the correct parton-jet assignments, where, in both cases, the same
events passing the $\Pgof>0.1$ requirement are used.

The $\DRbb > 2.0$ and $\Pgof>0.1$ requirements greatly reduce the background from QCD multijet production
from approximately 80 to 25\%,
but a significant number of multijet events enters the signal selection owing to the large production cross section
of that background contribution.
These events can fulfill the goodness-of-fit criterion
because of combinatorial chance, but not because of an underlying decay topology.
Therefore, it is assumed that \cPqb~jets can be exchanged with light-flavor jets for
the estimation of the background from data, because the probability for mimicking the
\ttbar topology is the same.

For the background estimation, the same selection as for the signal is applied, as described above,
but instead of requiring two \cPqb-tagged jets, events with exactly zero \cPqb-tagged jets are used.
For this veto, a very loose working point is used for the \cPqb~tagger,
to suppress contamination from \ttbar~events in this QCD-enriched sample.
A prescaled trigger similar to the signal trigger is used for this selection, which does not require
the presence of \cPqb~jets.
The kinematic fit is applied as before, but here any of the six light-flavor jets can be assigned to
the partons originating from the $\PW$ decays, as well as to the partons serving as \cPqb~quarks,
leading to 90 possible permutations that have to be evaluated.
This method allows one to determine the kinematic distributions of the background, but the normalization
is unknown.
In all plots, the background is normalized to the difference of the number of
data events and the number of expected signal events.
This data sample contains approximately five times
the number of expected background events, so it provides good statistical precision.

\begin{figure*}[htb]
  \includegraphics[width=.5\linewidth]{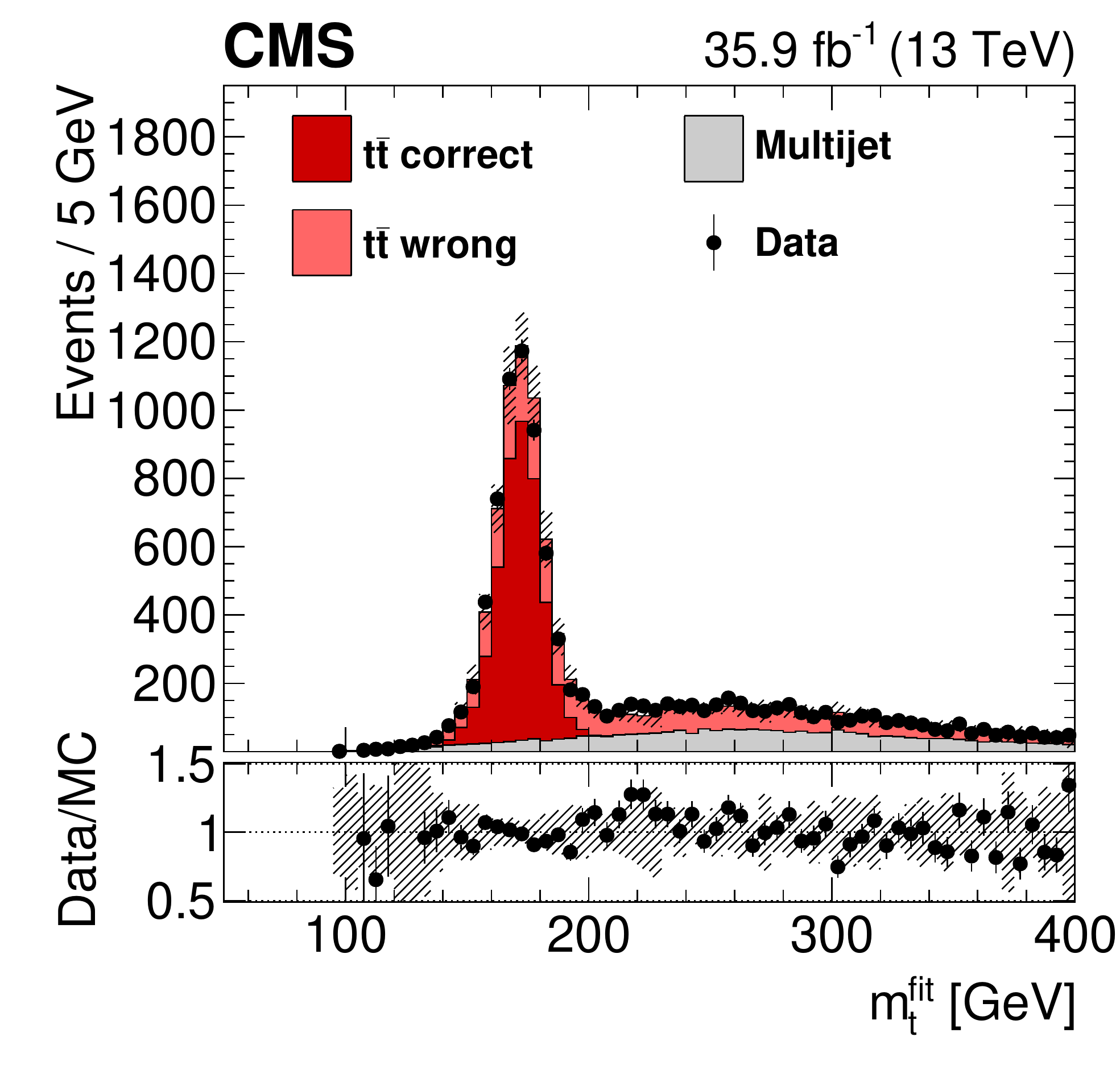}%
  \includegraphics[width=.5\linewidth]{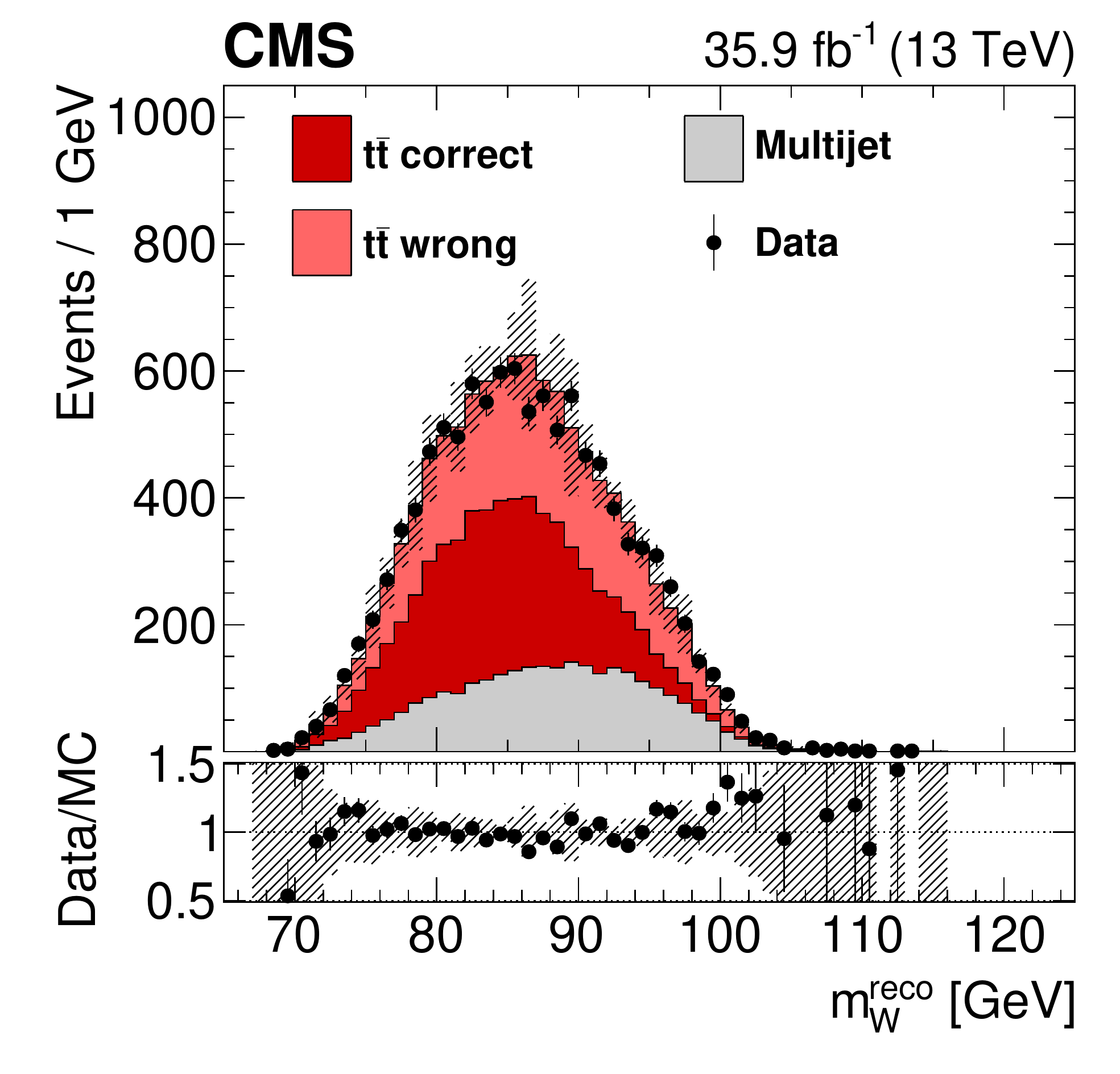}%
  \caption{The fitted top quark mass (left) and reconstructed \PW~boson mass (right) distributions of data
    compared to simulated signal and the multijet background estimate.
    The shown reconstructed \PW~boson mass is the average mass of the two W bosons in the event.
    For each event, the parton-jet assignment yielding the smallest \chisq in the kinematic fit is used.
    The simulated signal events are classified as correct or wrong assignments
    and displayed separately, and the distributions are normalized to the integrated luminosity.
    For the background estimate, the total normalization is given by the difference of observed data events and
    expected signal events.
    The hashed bands represent the total uncertainty in the prediction.
    The lower panels show the ratio of data and prediction.
    \label{fig:masses}}
\end{figure*}

The final selected data set consists of 10\,799 events with a signal purity of $75\%$.
Figure~\ref{fig:cp} shows the distributions of the separation of the two \cPqb~jets \DRbb and the quantity \Pgof in data, compared to the background estimate and \ttbar simulation.
For the \ttbar~signal, correct and wrong parton-jet assignments are shown separately.
The corresponding distributions of \mtfit and the reconstructed \PW~boson mass \mWreco, calculated from the originally reconstructed jets, are shown in Fig.~\ref{fig:masses}.
These two quantities are used in the top quark mass extraction described in the following section.

\section{Ideogram method}

For the extraction of \mtop, the ideogram method is used~\cite{Abdallah:2008xh,Chatrchyan:2012cz}.
Simultaneously, a JSF is determined that is used in addition to the standard CMS
jet energy calibration~\cite{Khachatryan:2015hba} to reduce the corresponding systematic uncertainty.
The distributions of \mtfit obtained from the kinematic fit and
\mWreco are used in a combined fit.
For \mWreco, the average mass of the two \PW~bosons in an event is used.

\begin{sloppypar}
The likelihood
\begin{linenomath*}\begin{align*}
\mathcal{L}\left(\mtop,\text{JSF}\right) &= P\left(\text{sample}|\mtop,\text{JSF}\right)\\
                                         &= \prod_{\text{events}}P\left(\text{event}|\mtop,\text{JSF}\right)\\
                                         &= \prod_{\text{events}}P\left(\mtfit,\mWreco|\mtop,\text{JSF}\right)
\end{align*}\end{linenomath*}
is maximized, yielding the best fit values for \mtop and JSF.
A prior probability for the \JSF can be incorporated by maximizing
\begin{linenomath*}\begin{equation*}
P(\JSF) P\left(\text{sample}|\mtop,\text{JSF}\right)
\end{equation*}\end{linenomath*}
instead.
Treating \mtfit and \mWreco as uncorrelated, as verified using simulated events, the probability $P\left(\mtfit,\mWreco|\mtop,\JSF\right)$ factorizes into
\begin{linenomath}
\ifthenelse{\boolean{cms@external}}
{
\begin{align*}
&P\left(\mtfit,\mWreco|\mtop,\text{JSF}\right)\\
&\quad{}={}f_\text{sig} P\left(\mtfit,\mWreco|\mtop,\text{JSF}\right)\\
&\qquad{}+{}\left(1-f_\text{sig}\right) P_\text{bkg}\left(\mtfit,\mWreco\right)\\
&\quad{}={}f_\text{sig} \sum_{j}f_{j}P_{j}\left(\mtfit|\mtop,\text{JSF}\right) P_{j}\left(\mWreco|\mtop,\text{JSF}\right)\\
&\qquad{}+{}\left(1-f_\text{sig}\right) P_\text{bkg}\left(\mtfit\right) P_\text{bkg}\left(\mWreco\right),
\end{align*}
}
{
\begin{align*}
P\left(\mtfit,\mWreco|\mtop,\text{JSF}\right) ={} & f_\text{sig} P\left(\mtfit,\mWreco|\mtop,\text{JSF}\right)\\
                                                  & {}+ \left(1-f_\text{sig}\right) P_\text{bkg}\left(\mtfit,\mWreco\right)\\
                                              ={} & f_\text{sig} \sum_{j}f_{j}P_{j}\left(\mtfit|\mtop,\text{JSF}\right) P_{j}\left(\mWreco|\mtop,\text{JSF}\right)\\
                                                  & {}+ \left(1-f_\text{sig}\right) P_\text{bkg}\left(\mtfit\right) P_\text{bkg}\left(\mWreco\right),
\end{align*}
}
\end{linenomath}
where $f_j$ with $j\in\left\{ \text{correct}, \text{wrong}\right\}$ is the relative fraction of the different permutation cases
and $f_\text{sig}$ is the signal fraction.
\end{sloppypar}

\begin{sloppypar}
The probability densities $P_{j}\left(\mtfit|\mtop,\text{JSF}\right)$ and $P_{j}\left(\mWreco|\mtop,\text{JSF}\right)$
for the signal are described by analytic functions parametrized in \mtop and JSF.
For the determination of the parameters, a simultaneous fit to simulated samples for seven different generated top quark
masses \mtgen and five different input JSF values is used.
The background shape is described by a spline interpolation as a function of \mtfit{} and \mWreco, but independent of the model parameters \mtop and JSF.
\end{sloppypar}

\begin{sloppypar}
Three variations of a maximum likelihood fit are performed to extract the top quark mass.
In the one-dimensional (1D) analysis, the JSF is fixed to unity (corresponding to a Dirac delta function for the prior probability),
\ie, the standard CMS jet energy calibration.
For the two-dimensional (2D) analysis, the JSF is a free parameter in the maximum likelihood fit, making possible a compensation of part of the systematic
uncertainties.
The signal fraction and correct permutation fraction are free parameters in both cases.
The third (hybrid) method is a weighted combination of both approaches, corresponding to a measurement with a Gaussian constraint on the
JSF around unity.
In the limit of an infinitely narrow JSF constraint, the hybrid method is identical to the 1D method, while for an infinitely
broad prior probability distribution, the 2D method is recovered.
The width of the Gaussian constraint in the hybrid method is optimized to yield the smallest total uncertainty.
\end{sloppypar}

\begin{figure}[!ht]
  \centering
  \includegraphics[width=.5\textwidth]{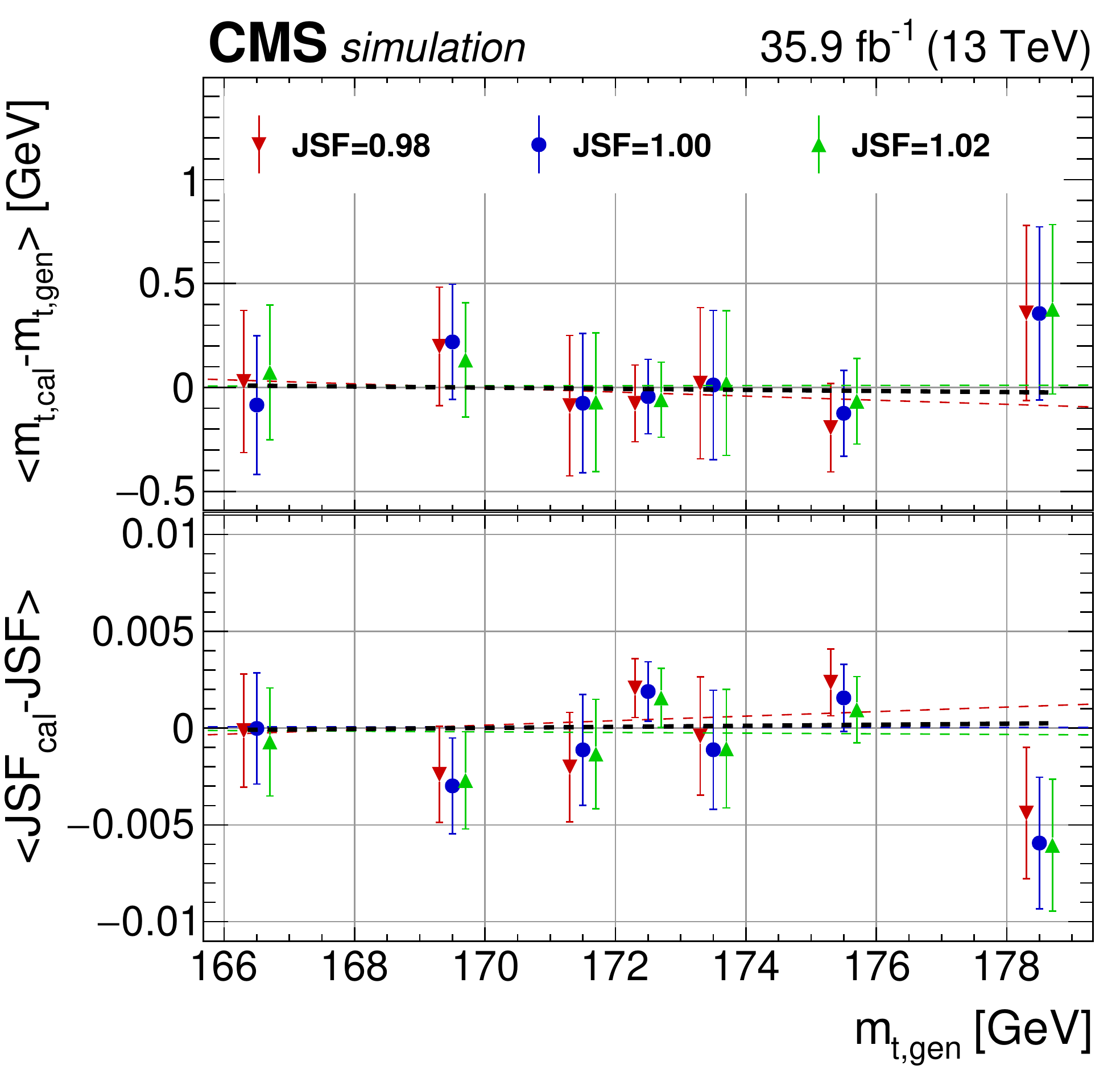}%
  \caption{Difference between extracted and generated top quark masses (upper panel) and JSFs (lower panel) for different input masses and JSFs
    after the calibration in the all-jets channel. The values are extracted using the 2D method.
  \label{fig:calibration}}
\end{figure}

\begin{sloppypar}
To calibrate the mass extraction method, pseudo-experiments are performed for the seven different generated values
of \mtgen and three input JSF values (0.98, 1.00, and 1.02).
The extracted \mtop and JSF values are compared to the input values and the residual slopes in \mtgen and JSF are used as calibration.
The residual biases after the calibration are shown in Fig.~\ref{fig:calibration} for pseudo-experiments
with different JSF and \mtgen values.
As expected, neither a significant residual offset nor a slope are observed after the calibration procedure.
\end{sloppypar}

\section{Systematic uncertainties}

A summary of the systematic uncertainty sources is shown in Table~\ref{tab:syst}.
The corresponding values are obtained from pseudo-experiments, using Monte Carlo (MC) signal samples with variations of the individual systematic uncertainty sources.
In the following, details for the determination of the most important uncertainties are given.
Most systematic uncertainty sources are shifted by $\pm 1$ standard deviation, and
the absolute value of the largest resulting shifts in $\mtop$ and JSF are quoted as systematic uncertainties
for the measurement.
For some uncertainties, different models are compared, and are described individually.
The maximum of the statistical uncertainty on the observed shift and the shift itself is used as the systematic uncertainty.

\begin{table*}[!htb]
  \topcaption{List of systematic uncertainties for the all-jets channel. The signs of the shifts ($\delta x = x_\text{variation} - x_\text{nominal}$) correspond to the $+1$ standard deviation variation of the systematic uncertainty source. For linear sums of the uncertainty groups, the relative signs have been considered.
    Shifts determined using dedicated samples for the systematic variation are displayed with the corresponding statistical uncertainty.
    \label{tab:syst}}
  \centering%
  \begin{tabular}{ld{2.7}d{2.1}d{2.7}d{2.7}d{2.1}}
    & \multicolumn{2}{c}{2D} & \multicolumn{1}{c}{1D} & \multicolumn{2}{c}{hybrid}\\
    & \multicolumn{1}{C}{\delta \mtop^{\text{2D}}}& \multicolumn{1}{C}{\delta \JSF^{\text{2D}}}& \multicolumn{1}{C}{\delta \mtop^{\text{1D}}}& \multicolumn{1}{C}{\delta \mtop^{\text{hyb}}}& \multicolumn{1}{C}{\delta\JSF^{\text{hyb}}}\\
    & \multicolumn{1}{c}{[GeV]}& \multicolumn{1}{c}{[\%]}& \multicolumn{1}{c}{[GeV]}& \multicolumn{1}{c}{[GeV]}& \multicolumn{1}{c}{[\%]}\\\hline
    \multicolumn{6}{l}{\textsl{Experimental uncertainties}}\\
    Method calibration                  &  0.06 &  0.2 &  0.06 &  0.06 &  0.2\\
    JEC (quad. sum)                     &  0.18 &  0.3 &  0.73 &  0.15 &  0.2\\
    -- Intercalibration                 & -0.04 & -0.1 & +0.12 & -0.04 & -0.1\\
    -- MPFInSitu                        & -0.03 &  0.0 & +0.22 & +0.08 & +0.1\\
    -- Uncorrelated                     & -0.17 & -0.3 & +0.69 & +0.12 & +0.2\\
    Jet energy resolution               & -0.09 & +0.2 & +0.09 & -0.04 & +0.1\\
    \cPqb~tagging                       &  0.02 &  0.0 &  0.01 &  0.02 &  0.0\\
    Pileup                              & -0.06 & +0.1 &  0.00 & -0.04 & +0.1\\
    Background                          &  0.10 &  0.1 &  0.03 &  0.07 &  0.1\\
    Trigger                             & +0.04 & -0.1 & -0.04 & +0.02 & -0.1\\
    [\cmsTabSkip]
    \multicolumn{6}{l}{\textsl{Modeling uncertainties}}\\
    JEC flavor (linear sum)             & -0.35 & +0.1 & -0.31 & -0.34 &  0.0\\
    -- light quarks (uds)               & +0.10 & -0.1 & -0.01 & +0.07 & -0.1\\
    -- charm                            & +0.03 &  0.0 & -0.01 & +0.02 &  0.0\\
    -- bottom                           & -0.29 &  0.0 & -0.29 & -0.29 &  0.0\\
    -- gluon                            & -0.19 & +0.2 & +0.03 & -0.13 & +0.2\\
    \cPqb~jet modeling (quad. sum)      &  0.09 &  0.0 &  0.09 &  0.09 &  0.0\\
    -- \cPqb~frag. Bowler--Lund         & -0.07 &  0.0 & -0.07 & -0.07 &  0.0\\
    -- \cPqb~frag. Peterson             & -0.05 &  0.0 & -0.04 & -0.05 &  0.0\\
    -- semileptonic \cPqb~hadron decays & -0.03 &  0.0 & -0.03 & -0.03 &  0.0\\
    PDF                                 &  0.01 &  0.0 &  0.01 &  0.01 &  0.0\\
    Ren. and fact. scales               &  0.05 &  0.0 &  0.04 &  0.04 &  0.0\\
    ME/PS matching                      & +0.32\pm0.20 & -0.3 & -0.05\pm0.14 & +0.24\pm0.18 & -0.2\\
    ISR PS scale                        & +0.17\pm0.17 & -0.2 & +0.13\pm0.12 & +0.12\pm0.14 & -0.1\\
    FSR PS scale                        & +0.22\pm0.12 & -0.2 & +0.11\pm0.08 & +0.18\pm0.11 & -0.1\\
    Top quark $\pt$                     & +0.03 &  0.0 & +0.02 & +0.03 &  0.0\\
    Underlying event                    & +0.16\pm0.19 & -0.3 & -0.07\pm0.14 & +0.10\pm0.17 & -0.2\\
    Early resonance decays              & +0.02\pm0.28 & +0.4 & +0.38\pm0.19 & +0.13\pm0.24 & +0.3\\
    CR modeling (max. shift)            & +0.41\pm0.29 & -0.4 & -0.43\pm0.20 & -0.36\pm0.25 & -0.3\\
    -- ``gluon move'' (ERD on)          & +0.41\pm0.29 & -0.4 & +0.10\pm0.20 & +0.32\pm0.25 & -0.3\\
    -- ``QCD inspired'' (ERD on)        & -0.32\pm0.29 & -0.1 & -0.43\pm0.20 & -0.36\pm0.25 & -0.1\\
    [\cmsTabSkip]
    Total systematic                    &  0.81 &  0.9 &  1.03 &  0.70 &  0.7\\
    Statistical (expected)              &  0.21 &  0.2 &  0.16 &  0.20 &  0.1\\
    Total (expected)                    &  0.83 &  0.9 &  1.04 &  0.72 &  0.7\\
  \end{tabular}
\end{table*}

\begin{itemize}
\item \textit{Method calibration:} The quadratic sum of the statistical uncertainty and the residual bias of the calibration curve (shown in Fig.~\ref{fig:calibration}) after the calibration is used as the systematic uncertainty.
\item \textit{JECs:} Jet energies are scaled up and down according to the \pt- and $\eta$-dependent
data/simulation uncertainties~\cite{Khachatryan:2016kdb}.
The correlation groups (called Intercalibration, MPFInSitu, and Uncorrelated) follow the recommendations documented in Ref.~\cite{CMS-PAS-JME-15-001}.
\item \textit{Jet energy resolution:} Since the jet energy resolution measured in data is worse than in simulation,
the simulation is modified to correct for the difference~\cite{Khachatryan:2016kdb}.
The jet energy resolution in the simulation is varied up and down within the uncertainty.
\item \textit{\cPqb~tagging:} The \pt-dependent uncertainty of the \cPqb~tagging efficiencies and misidentification rates of the CSVv2 \cPqb~tagger~\cite{Sirunyan:2017ezt} are taken into account by reweighting the simulated events accordingly.
\item \textit{Pileup:} To estimate the uncertainty in the determination of the number of pileup events and the reweighting procedure, the inelastic proton-proton cross section~\cite{Sirunyan:2018nqx} used in the determination is varied by ${\pm}4.6\%$.
\item \textit{Background:} An uncertainty in the background prediction is obtained by applying the method to simulation and comparing
the obtained estimate to the direct simulation, \ie, generated QCD multijet events passing the signal selection.
A linear fit to the ratio is consistent with a constant value of unity.
The slope is varied up and down within its uncertainty and used to reweight the events used for the determination of the background probability density function.
\item \textit{Trigger:} To estimate the uncertainty in the trigger selection, the data/simulation scale factor described
  in Section~\ref{sec:evtsel} is omitted. Additionally, a base trigger requiring the presence of one muon is used to obtain the correction factor.
  The maximum of the observed shifts with respect to the nominal correction is quoted as an uncertainty.

\item \textit{JEC flavor:} The difference between Lund string fragmentation and cluster fragmentation is evaluated
  comparing \PYTHIA~6.422~\cite{Sjostrand:2006za} and \HERWIGpp~2.4~\cite{Bahr:2008pv}.
  The jet energy response is compared separately for each jet flavor~\cite{Khachatryan:2016kdb}.
  Uncertainties for jets from different quark flavors and gluons are added linearly, which takes into account
  possible differences between the measured JSF, which is mainly sensitive to light quarks and gluons,
  and the \PQb{} jet energy scale.
\item \textit{\cPqb~jet modeling:} The uncertainty associated with the fragmentation of \cPqb~quarks is split into three components.
  The Bowler--Lund fragmentation function is varied within its uncertainties as determined by the ALEPH and DELPHI
  Collaborations~\cite{DELPHI:2011aa,Heister:2001jg}.
  As an alternative model of the fragmentation into \PQb hadrons, the Peterson fragmentation function is used and the difference
  obtained relative to the Bowler--Lund fragmentation function is assigned as an uncertainty.
  The third uncertainty source taken into account is the semileptonic \PQb hadron branching fraction, which is varied by
  $-0.45\%$ and $+0.77\%$, motivated by measurements of \PBz/\PBp decays and their corresponding uncertainties~\cite{Patrignani:2016xqp}.

\item \textit{PDF:}
  The 100 PDF replicas of the NNPDF3.0 NLO ($\alpS = 0.118)$ set are used to repeat the analysis~\cite{Ball:2014uwa}.
  The variance of the results is used to determine the PDF uncertainty.
  In addition, the $\alpS$ value is changed to 0.117 and 0.119.
  The maximum of the PDF uncertainty and the $\alpS$ variations is quoted as uncertainty.
\item \textit{Renormalization and factorization scales:}
  The renormalization and factorization scales for the ME calculation are varied.
  Both are multiplied independently from each other, and simultaneously by factors of 0.5 and 2 with respect to
  the default values.
  This is achieved by appropriately reweighting simulated events.
  The quoted uncertainty corresponds to the envelope of the resulting shifts.

\item \textit{ME/PS matching:} The matching of the \POWHEG{} ME calculations to the \PYTHIA PS is varied by shifting
  the parameter $h_{\text{damp}}=1.58^{+0.66}_{-0.59}$~\cite{CMS-PAS-TOP-16-021} within the uncertainties.
  The jet response $\pt^{\text{reco}}/\pt^{\text{gen}}$ as a function of $\pt^{\text{gen}}$ is rescaled in the variation samples to
  reproduce the response observed in the default sample.
\item \textit{ISR PS scale:} For initial-state radiation (ISR),
  the PS scale is varied in \PYTHIA.
  The ISR PS scale is multiplied by factors of 2 and 0.5 in dedicated MC samples.
\item \textit{FSR PS scale:}
  The PS scale used for final-state radiation (FSR) is scaled up by $\sqrt{2}$ and down by $1/\sqrt{2}$~\cite{Skands:2014pea}, affecting
  the fragmentation and hadronization, as well additional jet emission.
  The jet response is rescaled in the variation samples to reproduce the response observed in the default sample.
\item \textit{Top quark \pt:} Recent calculations suggest that the top quark \pt spectrum is strongly affected by next-to-next-to-leading-order effects~\cite{Czakon:2015owf}. The \pt of the top quark in simulation is varied to match the distribution measured by CMS~\cite{Khachatryan:2016mnb, Sirunyan:2017mzl} and its impact on the \mtop measurement is quoted as a systematic uncertainty.

\item \textit{Underlying event:} Measurements of the underlying event have been used to tune \PYTHIA parameters describing
  nonperturbative QCD effects~\cite{Skands:2014pea, CMS-PAS-TOP-16-021}.
  The parameters of the tune are varied within their uncertainties.
\item \textit{Early resonance decays:}
  Modeling of color reconnection (CR) introduces systematic uncertainties which are estimated by comparing different CR models
  and settings.
  In the default sample, the top quark decay products are not included in the CR process.
  This setting is compared to the case of including the decay products by enabling early resonance decays (ERD) in \PYTHIA8.
\item \textit{CR modeling:}
  In addition to the default model used in \mbox{\PYTHIA8}, two alternative CR models are used, namely
  a model with string formation beyond leading color (``QCD inspired'')~\cite{Christiansen:2015yqa}
  and a model allowing the gluons to be moved to another string (``gluon move'')~\cite{Argyropoulos:2014zoa}.
  Underlying event measurements are used to tune the parameters of all models~\cite{Skands:2014pea, CMS-PAS-TOP-16-021}.
  The largest shifts induced by the variations are assigned as the CR uncertainty.

  This approach, as well as the ERD variation, is new relative to the Run~1 results at $\sqrt{s}=7$ and $8\TeV$,
  because these CR models have become only recently available in \PYTHIA8.
  The new models were first used to evaluate the \mtop uncertainty due to CR in
  Ref.~\cite{Sirunyan:2018gqx}.
  Like in this analysis, the same increase in systematic uncertainty with respect to the Run~1 result has been observed.
\end{itemize}

A summary of the systematic uncertainties described above is given in Table~\ref{tab:syst}.
In Ref.~\cite{Sirunyan:2018gqx}, an ME generator uncertainty has been considered:
Instead of using \POWHEG~v2 as ME generator, the
\mbox{\MGvATNLO} 2.2.2 generator with the \texttt{FxFx} matching scheme is
used~\cite{Alwall:2014hca,Frederix:2012ps}.
The difference between the results obtained with the two generators is $\delta \mtop^{\text{hyb}}=+0.31\pm0.52$
for the hybrid method in the all-jets channel.
However, this is not significant because of the insufficient statistical precision of the available \MGvATNLO sample.
Since the radiation after the top quark decay is described by \PYTHIA, no significant impact of the ME generator choice
is expected beyond the variation of the PS scales and matching.
Therefore, no ME generator uncertainty is considered in the total uncertainty of the measurement,
but the number is just quoted here as a cross-check.

\section{Results}
For the 2D fit using the 10\,799 $\ttbar$ all-jets candidate events, the extracted parameters are
\begin{linenomath*}\begin{align*}
  \mtop^{\text{2D}} & = 172.43\pm0.22\statJSF\pm0.81\syst\GeV \text{ and}\\
  \JSF^{\text{2D}}  & = 0.996\pm0.002\stat\pm0.009\syst .
\end{align*}\end{linenomath*}
The corresponding 1D and hybrid fits yield instead
\begin{linenomath*}\begin{align*}
  \mtop^{\text{1D}} & = 172.13\pm0.17\stat\pm1.03\syst\GeV ,\\
  \mtop^{\text{hyb}} & = 172.34\pm0.20\statJSF\pm0.70\syst\GeV\text{, and}\\
  \JSF^{\text{hyb}}  & = 0.997\pm0.002\stat\pm0.007\syst .
\end{align*}\end{linenomath*}
In all cases the fitted values for the fraction of correct assignments, as well as the background fraction,
are in agreement with the values expected from simulation.
The hybrid measurement of $172.34\pm0.20\statJSF\pm0.43\CRERD\pm0.55\syst\GeV$
is the main result of this analysis, since it is constructed to provide the smallest uncertainty.
The color reconnection and early resonance decay parts are separated from the rest of the systematic uncertainties.
Because of the larger data sample used in this analysis, the statistical uncertainty is reduced with respect to the result
of $\mtop=172.32\pm0.25\statJSF\pm0.59\syst\GeV$ obtained at $\sqrt{s}=8\TeV$.
The new result is in good agreement with the value measured at $\sqrt{s}=8\TeV$,
where a leading-order \ttbar simulation has been employed to calibrate
the measurement, whereas an NLO simulation has been used here.
The systematic uncertainty is increased with respect to the Run~1 result,
because a broader set of CR models has been compared, which have become available in \PYTHIA~8.

\section{Combined measurement with the lepton+jets final state}

This measurement is combined with the lepton+jets final state, where only electrons and muons
are explicitly considered as leptons, while tau leptons enter the selection only when they decay leptonically.
The corresponding analysis for the lepton+jets final state is described in Ref.~\cite{Sirunyan:2018gqx}.
All selection and analysis steps are kept unchanged.
Since the same method for the mass extraction is used, a combination with the all-jets channel at the likelihood level is possible.

The total likelihood \LH is constructed from the single-channel likelihoods $\LH_i$,
\begin{linenomath*}\begin{equation*}\label{eq:naive_comb}
\LH(\mtop, \JSF) = \LHA(\mtop, \JSF) \LHL(\mtop, \JSF) ,
\end{equation*}\end{linenomath*}
where the indices A and L indicate the all-jets and lepton+jets channel, respectively.

\begin{figure}[!ht]
  \centering
  \includegraphics[width=.5\textwidth]{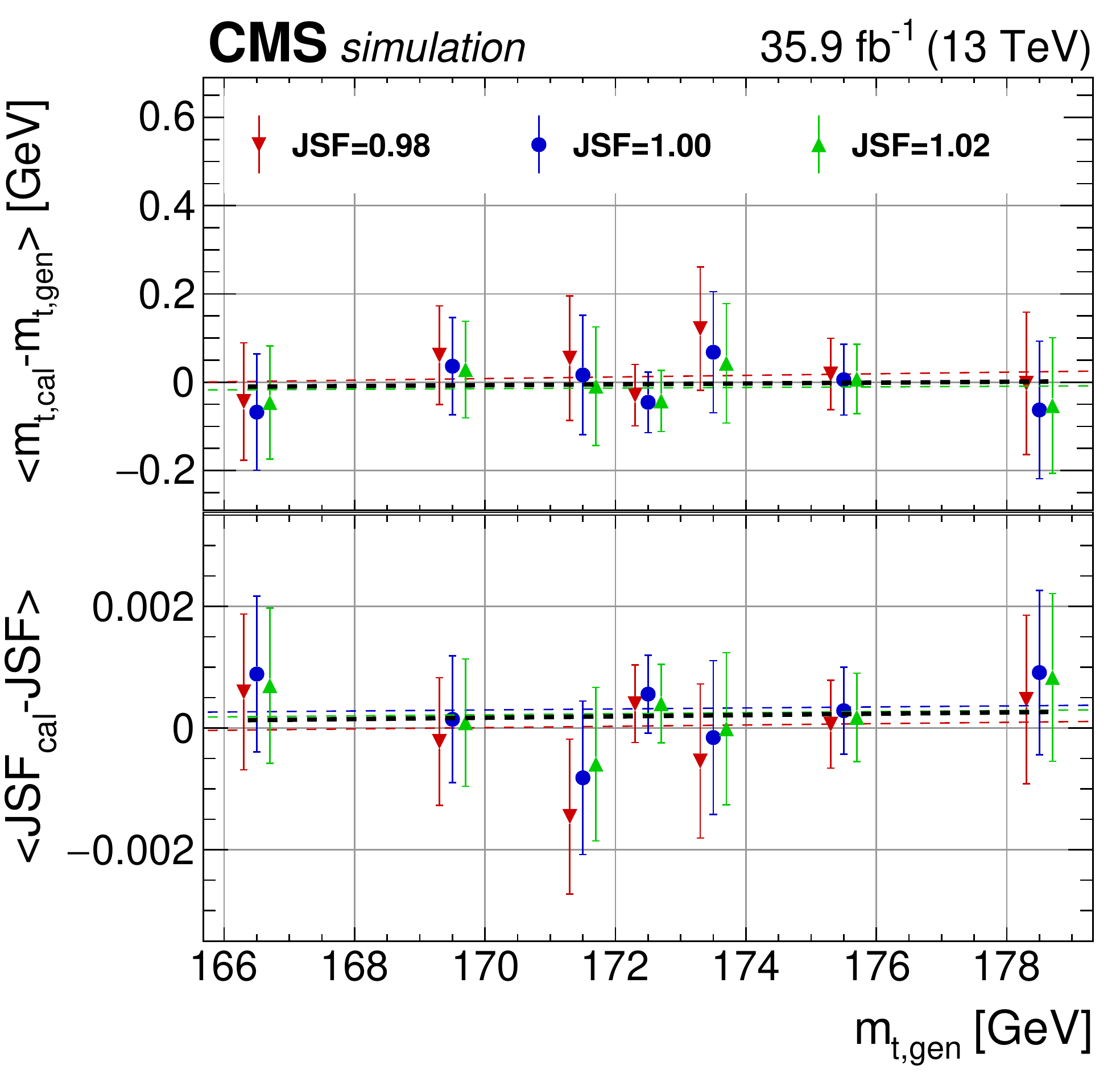}%
  \caption{Difference between extracted and generated top quark masses (upper panel) and JSFs (lower panel) for different input masses and JSFs
    after the single-channel calibrations for the combined measurement. The values are extracted using the 2D method.
  \label{fig:comb_calibration}}
\end{figure}

No extra calibration of the mass extraction is performed, but the single-channel calibrations are applied.
Figure~\ref{fig:comb_calibration} shows the extracted values for the top quark mass and JSF for different input values as a validation.
No residual dependence is observed.

The systematic uncertainties are evaluated as described above for the all-jets channel.
For the pseudo-experiments, the systematic uncertainty sources are varied simultaneously for both channels.
An exception are uncertainties that only affect a single channel.
These uncertainty sources are only varied for the corresponding channel.
For the all-jets channel, these are the background and trigger uncertainties.
In addition, uncertainties specific to the lepton+jets channel are introduced,
including the background and trigger uncertainties, as well as the uncertainties arising from the lepton isolation and
identification criteria, and are described in Ref.~\cite{Sirunyan:2018gqx}.
The complete list of uncertainties is shown in Table~\ref{tab:syst_comb}.
A comparison of the hybrid mass uncertainties can be found in Table~\ref{tab:syst_comp} for the all-jets and lepton+jets channels
as well as for the combination.
In general, the uncertainties for the combination are smaller than those for the all-jets channel and are close to the lepton+jets uncertainties, as expected because the combination is dominated by this channel.
The total uncertainty for the combination is slightly smaller than that for the lepton+jets channel.

\begin{table*}[!htb]
  \topcaption{List of systematic uncertainties for the combined mass extraction. The signs of the shifts ($\delta x = x_\text{variation} - x_\text{nominal}$) correspond to the $+1$ standard deviation variation of the systematic uncertainty source. For linear sums of the uncertainty groups, the relative signs have been considered.
    Shifts determined using dedicated samples for the systematic variation are displayed with the corresponding statistical uncertainty.
    \label{tab:syst_comb}}
  \centering%
  \begin{tabular}{ld{2.7}d{2.1}d{2.7}d{2.7}d{2.1}}
    & \multicolumn{2}{c}{2D} & \multicolumn{1}{c}{1D} & \multicolumn{2}{c}{hybrid}\\
    & \multicolumn{1}{C}{\delta \mtop^{\text{2D}}}& \multicolumn{1}{C}{\delta \JSF^{\text{2D}}}& \multicolumn{1}{C}{\delta \mtop^{\text{1D}}}& \multicolumn{1}{C}{\delta \mtop^{\text{hyb}}}& \multicolumn{1}{C}{\delta\JSF^{\text{hyb}}}\\
    & \multicolumn{1}{c}{[GeV]}& \multicolumn{1}{c}{[\%]}& \multicolumn{1}{c}{[GeV]}& \multicolumn{1}{c}{[GeV]}& \multicolumn{1}{c}{[\%]}\\\hline
    \multicolumn{6}{l}{\textsl{Experimental uncertainties}}\\
    Method calibration                  &  0.03 &  0.0 &  0.03 &  0.03 &  0.0\\
    JEC (quad. sum)                     &  0.12 &  0.2 &  0.82 &  0.17 &  0.3\\
    -- Intercalibration                 & -0.01 &  0.0 & +0.16 & +0.04 & +0.1\\
    -- MPFInSitu                        & -0.01 &  0.0 & +0.23 & +0.07 & +0.1\\
    -- Uncorrelated                     & -0.12 & -0.2 & +0.77 & +0.15 & +0.3\\
    Jet energy resolution               & -0.18 & +0.3 & +0.09 & -0.10 & +0.2\\
    \cPqb~tagging                       &  0.03 &  0.0 &  0.01 &  0.02 &  0.0\\
    Pileup                              & -0.07 & +0.1 & +0.02 & -0.05 & +0.1\\
    All-jets background                 &  0.01 &  0.0 &  0.00 &  0.01 &  0.0\\
    All-jets trigger                    & +0.01 &  0.0 &  0.00 & +0.01 &  0.0\\
    $\ell$+jets Background              & -0.02 &  0.0 & +0.01 & -0.01 &  0.0\\
    $\ell$+jets Trigger                 &  0.00 &  0.0 &  0.00 &  0.00 &  0.0\\
    Lepton isolation                    &  0.00 &  0.0 &  0.00 &  0.00 &  0.0\\
    Lepton identification               &  0.00 &  0.0 &  0.00 &  0.00 &  0.0\\
    [\cmsTabSkip]
    \multicolumn{6}{l}{\textsl{Modeling uncertainties}}\\
    JEC flavor (linear sum)             & -0.39 & +0.1 & -0.31 & -0.37 & +0.1\\
    -- light quarks (uds)               & +0.11 & -0.1 & -0.01 & +0.07 & -0.1\\
    -- charm                            & +0.03 &  0.0 & -0.01 & +0.02 &  0.0\\
    -- bottom                           & -0.31 &  0.0 & -0.31 & -0.31 &  0.0\\
    -- gluon                            & -0.22 & +0.3 & +0.02 & -0.15 & +0.2\\
    \cPqb~jet modeling (quad. sum)      &  0.08 &  0.1 &  0.04 &  0.06 &  0.1\\
    -- \cPqb~frag. Bowler--Lund         & -0.06 & +0.1 & -0.01 & -0.05 &  0.0\\
    -- \cPqb~frag. Peterson             & -0.03 &  0.0 &  0.00 & -0.02 &  0.0\\
    -- semileptonic \cPqb~hadron decays & -0.04 &  0.0 & -0.04 & -0.04 &  0.0\\
    PDF                                 &  0.01 &  0.0 &  0.01 &  0.01 &  0.0\\
    Ren. and fact. scales               &  0.01 &  0.0 &  0.02 &  0.01 &  0.0\\
    ME/PS matching                      & -0.10\pm0.08 & +0.1 & +0.02\pm0.05 & +0.07\pm0.07 & +0.1\\
    ME generator                        & +0.16\pm0.21 & +0.2 & +0.32\pm0.13 & +0.21\pm0.18 & +0.1\\
    ISR PS scale                        & +0.07\pm0.08 & +0.1 & +0.10\pm0.05 & +0.07\pm0.07 &  0.1\\
    FSR PS scale                        & +0.23\pm0.07 & -0.4 & -0.19\pm0.04 & +0.12\pm0.06 & -0.3\\
    Top quark $\pt$                     & +0.01 & -0.1 & -0.06 & -0.01 & -0.1\\
    Underlying event                    & -0.06\pm0.07 & +0.1 & +0.00\pm0.05 & -0.04\pm0.06 & +0.1\\
    Early resonance decays              & -0.20\pm0.08 & +0.7 & +0.42\pm0.05 & -0.01\pm0.07 & +0.5\\
    CR modeling (max. shift)            & +0.37\pm0.09 & -0.2 & +0.22\pm0.06 & +0.33\pm0.07 & -0.1\\
    -- ``gluon move'' (ERD on)          & +0.37\pm0.09 & -0.2 & +0.22\pm0.06 & +0.33\pm0.07 & -0.1\\
    -- ``QCD inspired'' (ERD on)        & -0.11\pm0.09 & -0.1 & -0.21\pm0.06 & -0.14\pm0.07 & -0.1\\
    [\cmsTabSkip]
    Total systematic                    &  0.71 &  1.0 &  1.07 &  0.61 &  0.7\\
    Statistical (expected)              &  0.08 &  0.1 &  0.05 &  0.07 &  0.1\\
    Total (expected)                    &  0.72 &  1.0 &  1.08 &  0.61 &  0.7\\
  \end{tabular}
\end{table*}

\begin{table*}[!htb]
  \topcaption{Comparison of the hybrid mass uncertainties for the all-jets and lepton+jets~\cite{Sirunyan:2018gqx} channels, as well as the combination.
    The signs of the shifts follow the convention of Tables~\ref{tab:syst} and~\ref{tab:syst_comb}.
    \label{tab:syst_comp}}
  \centering%
  \begin{tabular}{ld{1.2}d{1.2}d{1.2}}
      & \multicolumn{3}{C}{\delta \mtop^{\text{hyb}}~\text{[GeV]}}\\
      &\multicolumn{1}{c}{all-jets}&\multicolumn{1}{c}{$\ell$+jets}&\multicolumn{1}{c}{combination}\\\hline
    \multicolumn{4}{l}{\textsl{Experimental uncertainties}}\\
    Method calibration                  &  0.06 & 0.05&  0.03 \\
    JEC (quad. sum)                     &  0.15 & 0.18&  0.17 \\
    -- Intercalibration                 & -0.04 &+0.04& +0.04 \\
    -- MPFInSitu                        & +0.08 &+0.07& +0.07 \\
    -- Uncorrelated                     & +0.12 &+0.16& +0.15 \\
    Jet energy resolution               & -0.04 &-0.12& -0.10 \\
    \cPqb~tagging                       &  0.02 & 0.03&  0.02 \\
    Pileup                              & -0.04 &-0.05& -0.05 \\
    All-jets background                 &  0.07 & -   &  0.01 \\
    All-jets trigger                    & +0.02 & -   & +0.01 \\
    $\ell$+jets background              &  -    &+0.02& -0.01 \\
    [\cmsTabSkip]
    \multicolumn{4}{l}{\textsl{Modeling uncertainties}}\\
    JEC flavor (linear sum)             & -0.34 &-0.39& -0.37 \\
    -- light quarks (uds)               & +0.07 &+0.06& +0.07 \\
    -- charm                            & +0.02 &+0.01& +0.02 \\
    -- bottom                           & -0.29 &-0.32& -0.31 \\
    -- gluon                            & -0.13 &-0.15& -0.15 \\
    \cPqb~jet modeling (quad. sum)      &  0.09 & 0.12&  0.06 \\
    -- \cPqb~frag. Bowler--Lund         & -0.07 &-0.05& -0.05 \\
    -- \cPqb~frag. Peterson             & -0.05 &+0.04& -0.02 \\
    -- semileptonic \cPqb~hadron decays & -0.03 &+0.10& -0.04 \\
    PDF                                 &  0.01 & 0.02&  0.01 \\
    Ren. and fact. scales               &  0.04 & 0.01&  0.01 \\
    ME/PS matching                      & +0.24 &-0.07& +0.07 \\
    ME generator                        & -     &+0.20& +0.21 \\
    ISR PS scale                        & +0.14 &+0.07& +0.07 \\
    FSR PS scale                        & +0.18 &+0.13& +0.12 \\
    Top quark $\pt$                     & +0.03 &-0.01& -0.01 \\
    Underlying event                    & +0.17 &-0.07& -0.06 \\
    Early resonance decays              & +0.24 &-0.07& -0.07 \\
    CR modeling (max. shift)            & -0.36 &+0.31& +0.33 \\
    -- ``gluon move'' (ERD on)          & +0.32 &+0.31& +0.33 \\
    -- ``QCD inspired'' (ERD on)        & -0.36 &-0.13& -0.14 \\
    [\cmsTabSkip]
    Total systematic                    &  0.70 & 0.62&  0.61 \\
    Statistical (expected)              &  0.20 & 0.08&  0.07 \\
    Total (expected)                    &  0.72 & 0.63&  0.61 \\
  \end{tabular}
\end{table*}

\begin{figure}[ht]
  \centering
    \includegraphics[width=\cmsFigWidth]{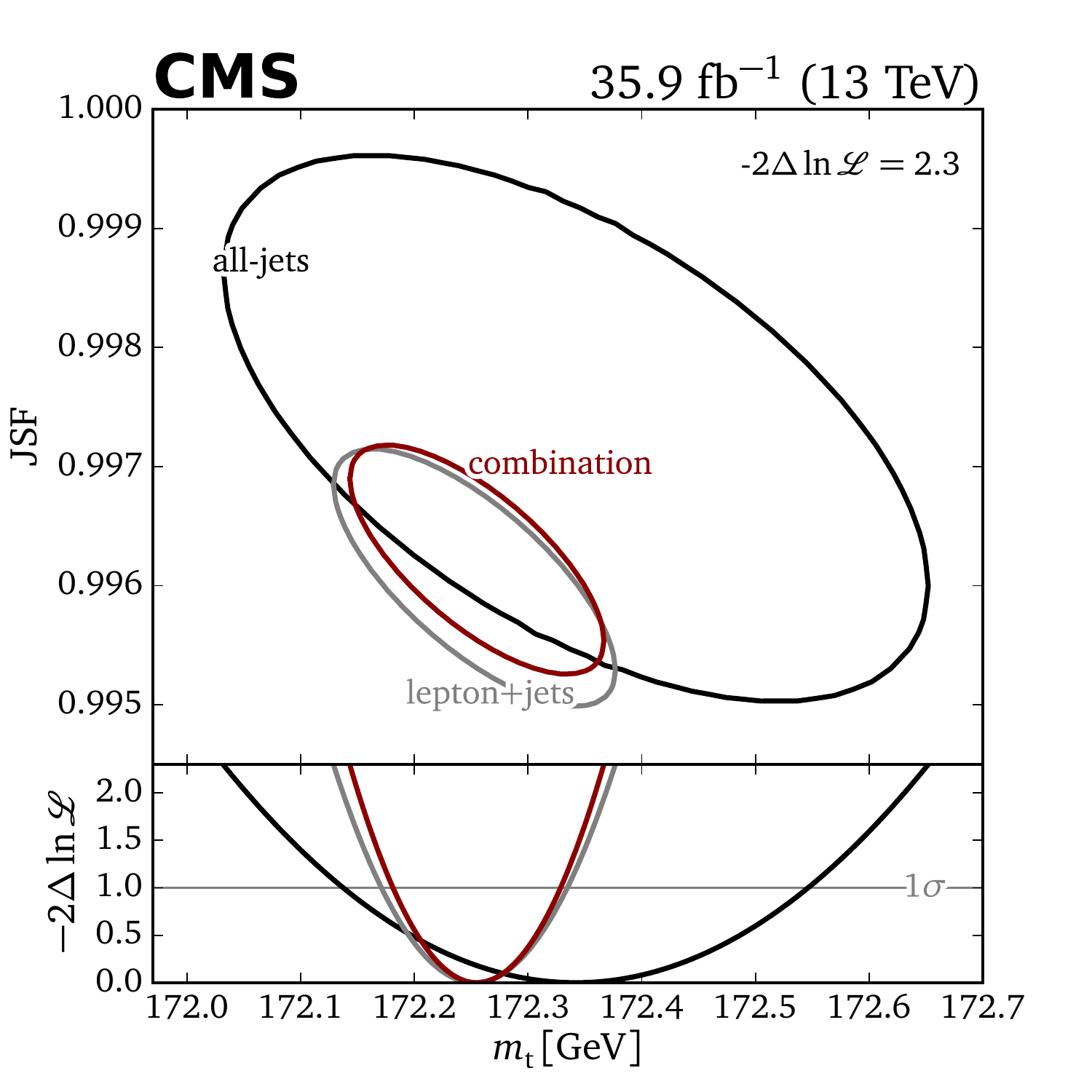}
  \caption{Likelihood contours for $-2\Delta\ln\LH=2.3$, corresponding to the $68\%$ confidence level, in the \mtop-\JSF plane (upper panel)
    and the likelihood profiles for the top quark mass (lower panel), where the level corresponding to one standard deviation ($\sigma$) is indicated.
    The hybrid measurement results for the all-jets and lepton+jets channels, as well as for the combination, are shown.
  \label{fig:contours_comb}}
\end{figure}

The combined measurement yields
\begin{linenomath*}\begin{align*}
  \mtop^{\text{2D}} & = 172.39\pm0.08\statJSF\pm0.71\syst\GeV \text{ and}\\
  \JSF^{\text{2D}}  & = 0.995\pm0.001\stat\pm0.010\syst
\end{align*}\end{linenomath*}
for the 2D method and
\begin{linenomath*}\begin{align*}
  \mtop^{\text{1D}} & = 171.94\pm0.05\stat\pm1.07\syst\GeV ,\\
  \mtop^{\text{hyb}} & = 172.26\pm0.07\statJSF\pm0.61\syst\GeV\text{, and}\\
  \JSF^{\text{hyb}}  & = 0.996\pm0.001\stat\pm0.007\syst
\end{align*}\end{linenomath*}
for the 1D and hybrid fits.
The likelihood contours for $-2\Delta\ln\LH=2.3$, corresponding to the $68\%$ confidence level, in the \mtop-\JSF plane are shown in Fig.~\ref{fig:contours_comb}
for the hybrid measurement results for the all-jets and lepton+jets channels, as well as for the combination.
Additionally, the likelihood profiles are displayed as a function of \mtop.
Both channels are in statistical agreement with each other.
The result of the combination is closer to the lepton+jets channel, as expected.
\ifthenelse{\boolean{cms@external}}
{}
{\clearpage}

Just as for the single-channel results, the hybrid measurement provides the best precision and is considered the main result.
This is the first top quark mass measurement using the $\ttbar$ lepton+jets and all-jets final states
combined in a single likelihood function.
The largest uncertainty contribution is related to the modeling of color reconnection, as it was observed for
the all-jets channel and the lepton+jets channel before using the same CR models.
Accordingly, the quoted systematic uncertainty is larger than those reported in the most precise combination
reported by the CMS Collaboration~\cite{Khachatryan:2015hba}, and comparable
to the value reported by the ATLAS Collaboration~\cite{Aaboud:2016igd}.

\section{Summary}
A measurement of the top quark mass (\mtop)  using the all-jets final state is presented.
The analyzed data set was collected by the CMS experiment in proton-proton collisions at $\sqrt{s}=13\,\TeV$
that correspond to an integrated luminosity of $35.9\fbinv$.
The kinematic properties in each event are reconstructed using a constrained fit that assumes a \ttbar{} hypothesis,
which suppresses the dominant multijet background and improves the mass resolution.

The value of \mtop and an additional jet energy scale factor (\JSF) are extracted using the ideogram method,
which uses the likelihood of the values of \mtop and \JSF
in each event to determine these parameters.
The resulting \mtop is measured to be
$172.34\pm0.20\statJSF\pm0.70\syst\GeV$.
This is in good agreement with previous CMS results obtained at
$\sqrt{s} = 7$, $8$, and $13\TeV$.
The modeling uncertainties are larger than in the previous measurements at lower
center-of-mass energies because of the use of new alternative color reconnection models that were not
previously available.

\begin{sloppypar}
A combined measurement using also the lepton+jets final state results in
$\mtop = 172.26\pm0.07\statJSF\pm0.61\syst\GeV$.
This is the first combined \mtop result obtained in the all-jets and lepton+jets final states
using a single likelihood function.
\end{sloppypar}

\begin{acknowledgments}
We congratulate our colleagues in the CERN accelerator departments for the excellent performance of the LHC and thank the technical and administrative staffs at CERN and at other CMS institutes for their contributions to the success of the CMS effort. In addition, we gratefully acknowledge the computing centers and personnel of the Worldwide LHC Computing Grid for delivering so effectively the computing infrastructure essential to our analyses. Finally, we acknowledge the enduring support for the construction and operation of the LHC and the CMS detector provided by the following funding agencies: BMBWF and FWF (Austria); FNRS and FWO (Belgium); CNPq, CAPES, FAPERJ, FAPERGS, and FAPESP (Brazil); MES (Bulgaria); CERN; CAS, MoST, and NSFC (China); COLCIENCIAS (Colombia); MSES and CSF (Croatia); RPF (Cyprus); SENESCYT (Ecuador); MoER, ERC IUT, and ERDF (Estonia); Academy of Finland, MEC, and HIP (Finland); CEA and CNRS/IN2P3 (France); BMBF, DFG, and HGF (Germany); GSRT (Greece); NKFIA (Hungary); DAE and DST (India); IPM (Iran); SFI (Ireland); INFN (Italy); MSIP and NRF (Republic of Korea); MES (Latvia); LAS (Lithuania); MOE and UM (Malaysia); BUAP, CINVESTAV, CONACYT, LNS, SEP, and UASLP-FAI (Mexico); MOS (Montenegro); MBIE (New Zealand); PAEC (Pakistan); MSHE and NSC (Poland); FCT (Portugal); JINR (Dubna); MON, RosAtom, RAS, RFBR, and NRC KI (Russia); MESTD (Serbia); SEIDI, CPAN, PCTI, and FEDER (Spain); MOSTR (Sri Lanka); Swiss Funding Agencies (Switzerland); MST (Taipei); ThEPCenter, IPST, STAR, and NSTDA (Thailand); TUBITAK and TAEK (Turkey); NASU and SFFR (Ukraine); STFC (United Kingdom); DOE and NSF (USA).

\hyphenation{Rachada-pisek} Individuals have received support from the Marie-Curie program and the European Research Council and Horizon 2020 Grant, contract No. 675440 (European Union); the Leventis Foundation; the A.P.\ Sloan Foundation; the Alexander von Humboldt Foundation; the Belgian Federal Science Policy Office; the Fonds pour la Formation \`a la Recherche dans l'Industrie et dans l'Agriculture (FRIA-Belgium); the Agentschap voor Innovatie door Wetenschap en Technologie (IWT-Belgium); the F.R.S.-FNRS and FWO (Belgium) under the ``Excellence of Science -- EOS" -- be.h project n.\ 30820817; the Ministry of Education, Youth and Sports (MEYS) of the Czech Republic; the Lend\"ulet (``Momentum") Programme and the J\'anos Bolyai Research Scholarship of the Hungarian Academy of Sciences, the New National Excellence Program \'UNKP, the NKFIA research grants 123842, 123959, 124845, 124850, and 125105 (Hungary); the Council of Science and Industrial Research, India; the HOMING PLUS program of the Foundation for Polish Science, cofinanced from European Union, Regional Development Fund, the Mobility Plus program of the Ministry of Science and Higher Education, the National Science Center (Poland), contracts Harmonia 2014/14/M/ST2/00428, Opus 2014/13/B/ST2/02543, 2014/15/B/ST2/03998, and 2015/19/B/ST2/02861, Sonata-bis 2012/07/E/ST2/01406; the National Priorities Research Program by Qatar National Research Fund; the Programa Estatal de Fomento de la Investigaci{\'o}n Cient{\'i}fica y T{\'e}cnica de Excelencia Mar\'{\i}a de Maeztu, grant MDM-2015-0509 and the Programa Severo Ochoa del Principado de Asturias; the Thalis and Aristeia programs cofinanced by EU-ESF and the Greek NSRF; the Rachadapisek Sompot Fund for Postdoctoral Fellowship, Chulalongkorn University and the Chulalongkorn Academic into Its 2nd Century Project Advancement Project (Thailand); the Welch Foundation, contract C-1845; and the Weston Havens Foundation (USA).
\end{acknowledgments}

\bibliography{auto_generated}

\cleardoublepage \appendix\section{The CMS Collaboration \label{app:collab}}\begin{sloppypar}\hyphenpenalty=5000\widowpenalty=500\clubpenalty=5000\vskip\cmsinstskip
\textbf{Yerevan Physics Institute, Yerevan, Armenia}\\*[0pt]
A.M.~Sirunyan, A.~Tumasyan
\vskip\cmsinstskip
\textbf{Institut f\"{u}r Hochenergiephysik, Wien, Austria}\\*[0pt]
W.~Adam, F.~Ambrogi, E.~Asilar, T.~Bergauer, J.~Brandstetter, M.~Dragicevic, J.~Er\"{o}, A.~Escalante~Del~Valle, M.~Flechl, R.~Fr\"{u}hwirth\cmsAuthorMark{1}, V.M.~Ghete, J.~Hrubec, M.~Jeitler\cmsAuthorMark{1}, N.~Krammer, I.~Kr\"{a}tschmer, D.~Liko, T.~Madlener, I.~Mikulec, N.~Rad, H.~Rohringer, J.~Schieck\cmsAuthorMark{1}, R.~Sch\"{o}fbeck, M.~Spanring, D.~Spitzbart, W.~Waltenberger, J.~Wittmann, C.-E.~Wulz\cmsAuthorMark{1}, M.~Zarucki
% ...
\vskip\cmsinstskip
% ...
1:  Also at Vienna University of Technology, Vienna, Austria\\
% ...
\end{sloppypar}
\end{document}